\documentclass{ws-ijmpcs}

\begin{document}

\markboth{A.G. Alaverdyan, G.B. Alaverdyan, Sh.R. Melikyan}
{Energy Release Associated with Quark Phase Transition in Neutron
Stars}

%%%%%%%%%%%%%%%%%%%%% Publisher's Area please ignore %%%%%%%%%%%%%%%
%
\catchline{}{}{}{}{}
%
%%%%%%%%%%%%%%%%%%%%%%%%%%%%%%%%%%%%%%%%%%%%%%%%%%%%%%%%%%%%%%%%%%%%

\title{ENERGY RELEASE ASSOCIATED WITH QUARK PHASE TRANSITION IN NEUTRON STARS: COMPARATIVE ANALYSIS OF MAXWELL AND GLENDENNING SCENARIOS}

\author{ANI ALAVERDYAN, GRIGOR ALAVERDYAN{\dag} and SHUSHAN MELIKYAN }

\address{Faculty of RadioPhysics, Yerevan State University, Manoogyan str.
1\\ Yerevan, 0025, Armenia\\{\dag} galaverdyan@ysu.am}

\maketitle

\begin{history}
\received{Day Month Year}
\revised{Day Month Year}
\end{history}

\begin{abstract}
We study the compact stars internal structure and observable
characteristics alterations due to the quark deconfinement phase
transition. To proceed with, we investigate the properties of
isospin- asymmetric nuclear matter in the improved relativistic
mean-field (RMF) theory, including a scalar-isovector
$\delta$-meson effective field. In order to describe the quark
phase, we use the improved version of the MIT bag model, in which
the interactions between $u$, $d $  and $s$ quarks inside the bag
are taken into account in the one-gluon exchange approximation. We
compute the amount of energy released by the corequake for both
cases of deconfinement phase transition scenarios, corresponding
to the Maxwellian type ordinary first-order phase transition and
the phase transition with formation of a mixed quark-hadron phase
(Glendenning scenario).

\end{abstract}

\ccode{PACS numbers: 97.60.Jd, 26.60.-c, 14.65.-q}

\section{Introduction}

Neutron stars are objects with a very complicated, many-component
and many-layered structure. The surface layer of such objects
consists of an ordinary matter with atomic and molecular
structure. In the inner layers of different depths, the conditions
both for the rearrangement of the structural formations and for
the creation of new constituents of matter are satisfied. In the
central region of neutron star the density of matter reaches such
high values that makes possible the appearance of various exotic
particle species and phases such as hyperons, deconfined $u$, $d$,
$s$ quarks and $\pi$, $K$ meson condensates. The existence of
compact stars consisting of matter in a deconfined quark phase was
predicted long ago\cite{Iv-Kurd}\cdash\cite{Bod}. Over the past
few decades many researchers had been intensively studied various
aspects related to the formation of exotic degrees of freedom in
neutron stars and clarification of the observationally testing of
dynamic processes, confirming the existence in the interiors of
stars such constituents (for review see, e.g.,
Refs.~\refcite{Gl_bk,Haen_bk} and references therein). Phase
transitions accompanied by discontinuities of the thermodynamic
potentials are the most interesting because it leads to a
dynamical rearrangement of neutron stars. Depending on the value
of surface tension $\sigma_{s}$, the phase transition of nuclear
matter into quark matter can occur in two scenarios\cite{Heis,Hj}:
either ordinary first order phase transition at constant pressure
with a density jump (Maxwell construction), or formation of mixed
hadron-quark matter with a continuous variation of pressure and
density (Glendenning construction)\cite{Gl1992}. The question of
whether the formation of a mixed phase is energetically favorable,
given the finite dimensions of the quark structures inside nuclear
matter, the Coulomb interaction, as well as the surface energy,
has been examined in Refs.~\refcite{Hj,Vos,Maruy}. It was shown
that the mixed phase is energetically favorable for small values
of the surface tension between the quark matter and the nuclear
matter. Uncertainty of the surface tension values makes it
impossible to determine the phase transition scenario which
actually takes place. An important manifestation of the
hadron-quark phase transition in the compact stars is a dynamical
process of accretion of matter onto the surface of neutron stars
with the hadron structure, which leads to fulfillment of
conditions for the formation in the center of the star a new phase
containing deconfined quarks. This type of transition can also
occur in case of a rotating neutron star that is slowing down when
the pressure in the center rises and exceeds the threshold value.

The process of catastrophic rearrangement with formation of a
quark core of finite radius at the star's center will be
accompanied by release of a colossal amount of energy comparable
to the energy release during a supernova explosion. These features
of accreting neutron star near the critical configuration make it
a potential candidate both for the gamma-ray bursts
(GRBs)\cite{Bom,Drag} and for the gravitational wave (GW) emission
sources\cite{Marang,Lin}. Note that a similar process of both
restructuring and energy release takes place also in the case of
pion condensation in the cores of neutron
stars\cite{Mig}\cdash\cite{Ber}.

Recent series of our articles\cite{Alav1}\cdash\cite{Alav5} were
devoted to a detailed investigation of quark deconfinement phase
transition of neutron star matter, when the nuclear matter is
described in the relativistic mean-field (RMF) theory with the
scalar-isovector $\delta$-meson effective field. The calculation
results of the mixed phase structure (Glendenning construction)
are compared with the results of a usual first-order phase
transition (Maxwell construction). This article is a continuation
of this series. Here we aim to investigate the energy release and
the change of integral parameters of compact stars due to the
quark phase transition in the two alternative scenarios and to
identify possible differences in the manifestations of this
phenomenon.

\section{Model Equations of State for Compact Stars}
We use the EOS of Baym-Bethe-Pethick (BBP)\cite{BBP} for
description of hadronic phase in the lower density region,
corresponding to the outer and inner crust of the star. In nuclear
and supranuclear density region ($n\geq 0.1$ fm$^{-3}$) was used
the relativistic Lagrangian density of many-particle system
consisting of nucleons, $p$, $n$, electrons and isoscalar-scalar
($\sigma$), isoscalar-vector ($\omega$), isovector-scalar
($\delta$), and isovector-vector ($\rho$) - exchanged mesons:

\begin{eqnarray}
\label{eq1} { \cal L}=\overline{\psi}_{N}[\gamma ^{\mu }(
i\partial _{\mu }-g_{\omega }\omega _{\mu }-\frac{1}{2}g_{\rho }%
\overrightarrow{\tau }_{N}\overrightarrow{\rho }_{\mu }) -(
m_{N}-g_{\sigma }\sigma
-g_{\delta}\overrightarrow{\tau}_{N}\overrightarrow{\delta })]
\psi _{N}\nonumber ~~\\+\frac{1}{2}( \partial_{\mu }\sigma
\partial ^{\mu }\sigma -m_{\sigma }\sigma ^{2})
-\frac{b}{3}~m_{N}(g_{\sigma}\sigma
)^{3}-\frac{c}{4}~(g_{\sigma}\sigma )^{4} \nonumber
~~\\+\frac{1}{2}m_{\omega }^{2}\omega ^{\mu }\omega _{\mu
}-\frac{1}{4}\Omega _{\mu \nu}\Omega ^{\mu \nu }%
+\frac{1}{2}m_{\rho }^{2}\overrightarrow{\rho }^{\mu }%
\overrightarrow{\rho }_{\mu }-\frac{1}{4}\Re_{\mu \nu }\Re^{\mu
\nu}\nonumber~~\\+\frac{1}{2}( \partial _{\mu
}\overrightarrow{\delta }\partial
^{\mu }\overrightarrow{\delta }-m_{\delta }^{2}\overrightarrow{%
\delta }^{2})+\overline{\psi} _{e}( {i\gamma ^{\mu }\partial
_{\mu}  - m_{e}})\,\psi _{e},~~
\end{eqnarray}

\noindent where $\sigma$, $\omega _{\mu }$,
$\overrightarrow{\delta }$, and $\overrightarrow{\rho }^{\mu }$
are the fields of the $\sigma$, $\omega$, $\delta$, and $\rho$
exchange mesons, respectively, $m_{N}$, $m_{e}$, $m_{\sigma}$,
$m_{\omega}$, $m_{\delta}$, $m_{\rho}$ are the masses of the free
particles, $\psi _{N} = \left({{\begin{array}{*{20}c} {\psi _{p}}  \hfill \\
{\psi _{n}} \hfill \\\end{array}} } \right)$ is the isospin
doublet for nucleonic bispinors, and $\overrightarrow{\tau}$ are
the isospin $2\times2$ Pauli matrices. Antisymmetric tensors of
the vector fields $\omega _{\mu }$ and $\overrightarrow{\rho}_{\mu
}$ given by
\begin{equation}
\label{eq3} \Omega _{\mu \nu} = \partial _{\mu} \omega _{\nu} -
\partial _{\nu}  \omega _{\mu} ,\quad \;\Re _{\mu \nu} =
\partial _{\mu}  \overrightarrow{\rho}_{\nu} -
\partial _{\nu}  \overrightarrow{\rho}_{\mu}.
\end{equation}

In our calculations we take
$a_{\delta}=\left(g_{\delta}/m_{\delta}\right)^2=2.5$ fm$^2$ for
the $\delta$ coupling constant, as in Ref.~\refcite{Liu}. Also we
use $m_{N}=938.93$ MeV for the bare nucleon mass,
$m_{N}^{\ast}=0.78~m_{N}$ for the nucleon effective mass,
$n_{0}=0.153$ fm$^{-3}$ for the baryon number density at
saturation, $f_{0}=-16.3$ MeV for the binding energy per baryon,
$K=300$ MeV for the incompressibility modulus, and
$E_{sym}^{(0)}=32.5$ MeV for the asymmetry energy. Then five other
constants, $a_{i}=\left(g_{i}/m_{i}\right)^2$ $(i=\sigma,~
\omega,~ \rho)$, $b$ and $c$, can be  numerically determined:
$a_{\sigma}=\left(g_{\sigma}/m_{\sigma}\right)^2=9.154$ fm$^2$,
$a_{\omega}=\left(g_{\omega}/m_{\omega}\right)^2=4.828$ fm$^2$,
$a_{\rho}=\left(g_{\rho}/m_{\rho}\right)^2=13.621$ fm$^2$,
$b=1.654\cdot10^{-2}$ fm$^{-1}$, $c=1.319\cdot10^{-2}$. The
knowledge of the model parameters makes it possible to solve the
set of four equations in a self-consistent way and to determine
the re-denoted mean-fields, $\sigma \equiv g_{\sigma}\bar
{\sigma}$, $\omega \equiv g_{\omega}\bar {\omega_{0}}$, $\delta
\equiv g_{\delta}\bar{\delta}^{\small(3)}$, and $\rho \equiv
g_{\rho}\bar {\rho_{0}}^{(3)}$, which depend on baryon number
density $n$ and asymmetry parameter $\alpha=(n_n-n_p)/n$. The
standard QHD procedure allows to obtain expressions for energy
density $\varepsilon(n,\alpha)$ and pressure $P(n,\alpha) $ (for
details see Ref.~\refcite{Alav1}).

To describe  the quark phase an improved version of the MIT bag
model was used, in which the interactions between $u,~d,~s$ quarks
inside the bag are taken in a one-gluon exchange
approximation\cite{Far}. We choose $m_{u} = 5$ MeV, $m_{d} = 7$
MeV and $m_{s} = 150$ MeV for quark masses,  $B=60$ MeV/fm$^3$ for
bag parameter and $\alpha_{s}=0.5$ for the strong interaction
constant.

\begin{figure}[pt]
\centerline{\psfig{file=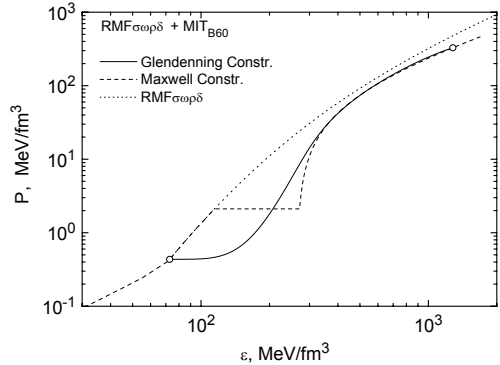,width=8 cm}} \vspace*{8pt}
\caption{EOS of neutron star matter for the two different
hadron-quark phase transition constructions. Solid and dashed
lines correspond to the Glendenning and Maxwell constructions,
respectively, and the dotted line to the pure $npe$ matter. Open
circles represent the mixed phase boundaries. \label{f1}}
\end{figure}

Using these equations of state for both the nucleonic and the
quark phase, we can calculate the physical parameters of the phase
transition as in case of Glendenning construction where these
phases satisfy the Gibbs condition, are electrically charged
separately, but lead to the global electrical neutrality of the
system, as well as in the case when both phases are separately
neutral and transition is a usual first-order phase transition
corresponding to the well-known Maxwell construction. Model EOSs
of neutron star matter for Glendenning and Maxwell construction
cases are presented in Fig.~\ref{f1}. In case of the Maxwell
construction the phase transition occurs at constant pressure
$P_0=2.11$~MeV/fm$^3$ and nucleonic matter of energy density
$\varepsilon_N=114.5$~MeV/fm$^3$ coexisted with the quark matter
of energy density $\varepsilon_Q=271.4$~MeV/fm$^3$. In Glendenning
construction case the deconfinement phase transition proceed
through formation of a mixed hadron-quark phase. Boundaries of the
mixed phase are $\varepsilon_N=72.79$~MeV/fm$^3$, $P_N=
0.43$~MeV/fm$^3$ and $\varepsilon_Q=1280.88$~MeV/fm$^3$, $P_Q=
327.75$~MeV/fm$^3$.

It was shown\cite{Seid} that for an ordinary first order phase
transition, the density discontinuity parameter $\lambda=
\varepsilon_{S} /(\varepsilon_N + P_0)$ plays a decisive in the
stability of neutron stars with arbitrarily small cores made of
the denser-phase matter. Here $P_0$ is coexistence pressure of two
phases, and $\varepsilon_N,~ \varepsilon_{S} $ are the energy
density of normal and superdense phases, respectively.
Paraphrasing the conclusions of that paper, in case of a
hadron-quark first order phase transition, we get the following
stability criterions: if $\lambda<3/2$, then the neutron star with
an arbitrarily small core of strange quark matter is stable, and
if $\lambda>3/2$, neutron stars with small quark cores are
unstable. In the latter case, for a stable star there is a nonzero
minimum value for the radius of the quark core. Note that in the
Maxwell construction case of deconfinement phase transition
discussed in this article the value of jump parameter is
$\lambda=2.327$.

\section{Changes in the Stellar Parameters and Energy Release due to Quark Phase Transition}
Using the neutron star matter EOSs obtained in previous section,
we have integrated the Tolman-Oppenheimer-Volkoff (TOV)
equations\cite{T,O-V} and obtained the gravitational mass $M$,
radius $R$ and baryonic mass $M_0=m_NN_B$ ($m_N$ is the nucleon
mass and $N_B$ the total number of baryons) of compact stars for
the different values of central pressure $P_c$.
\begin{figure}[pb]
\centerline{\psfig{file=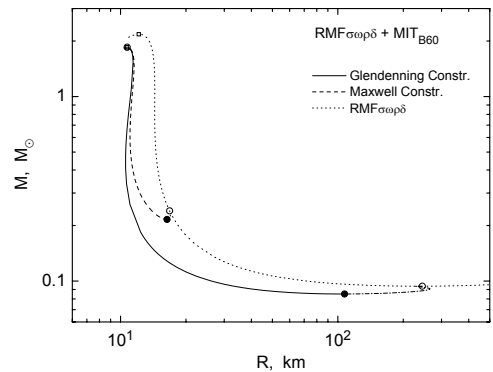,width=8 cm}} \vspace*{8pt}
\caption{Mass-radius relations for model EOSs of neutron star
matter presented in Fig.~\ref{f1}. The meaning of the curves is
the same as in Fig.~\ref{f1}. Open circles denote the critical
configurations, solid circles denote the stable hybrid stars with
minimal mass. Dash-dotted line between circles denote unstable
stars branch. \label{f2}}
\end{figure}

In Fig.~\ref{f2} we show the $M(R)$ dependence of compact stars
for different quark deconfinement phase transition scenarios. Open
circles denote the critical configurations while solid circles
denote the stable stars with minimal mass  in the center of which
there are deconfined quarks. Moreover, in the case of Maxwell
construction this core is consisted of a quark-electron plasma,
while in the Glendenning construction case it is consisted of a
mixed quark-hadron matter. We can see that in both cases of phase
transition scenarios considered here there are unstable star
branches between critical and minimum-mass configurations
(dot-dashed line segment). The fact that in the case of Maxwell
construction the  $M-R$ relation has such a behavior is not
surprising, since according to the Seidov criterium\cite{Seid} the
infinitesimal core of denser matter in ordinary first order phase
transition is unstable when $\lambda>3/2$. The corresponding EOS
of neutron star matter considered here satisfies this condition.
With regard to the case of Glendenning construction, the
appearance of the unstable branch of compact stars with
infinitesimal core consisting of a mixed hadron-quark matter, is
not the standard case. The fact that in this scenario, the energy
density is a continuous function of pressure and in most cases
leads to a monotonic increase of stars mass at the lower boundary
of the mixed phase, means that the configurations with an
infinitesimal core of the mixed phase in many cases are stable.
The case of the equation of state with a mixed phase, leading to a
similar behavior of the stellar mass as a function of the central
pressure was examined in the article\cite{Mish}. In contrast to
our case, in this article the unstable branch of compact stars
appears near the upper threshold of the mixed phase and
catastrophic transition from hybrid star  with a smaller core of
mixed phase to the hybrid star with a large core of the mixed
phase is realized. In case of Glendenning construction considered
here the branch of unstable compact stars appear near the lower
threshold of the mixed phase. In this case the accretion of matter
onto the surface of an ordinary neutron star located below the
critical configuration, the baryonic mass of the star increases,
the star is reached a critical configuration and the transition to
the configuration with deconfined quark  phase is takes place.

It is worth noting that the maximum mass of stars, containing
deconfined quarks   is $M_{max}=1.853 M_\odot$ for the case of
Glendenning construction and $M_{max}=1.828 M_\odot$ for the case
of Maxwell construction.

\begin{figure}[pb]
\centerline{\psfig{file=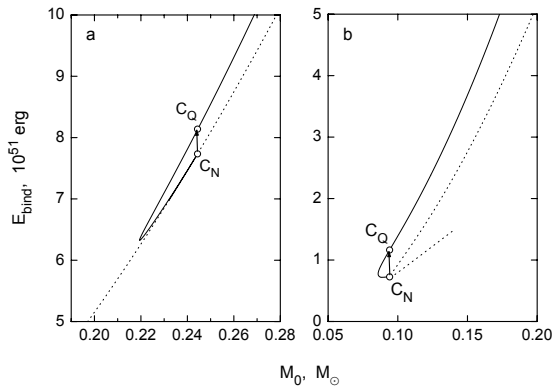,width=8 cm}} \vspace*{8pt}
\caption{Binding energy of compact stars as a function of baryonic
mass $M_0$ for Maxwell (left panel) and Glendenning (right panel)
constructions. Dotted line corresponds to the pure nuclear matter.
$C_N$ denotes the critical configuration of hadronic star, and
$C_Q$ presents the neutron star with same baryonic mass containing
quark phase. \label{f3}}
\end{figure}
Since the transition of ordinary neutron star to a star containing
quark matter occurs at a constant baryon number, then the
characteristics of the star is conveniently to represent as
functions of baryonic mass $M_0$. The binding energy $E_{bind}
=(M-M_0)c^2$ of compact stars as a function of baryonic mass $M_0$
for Maxwell and Glendenning hadron-quark phase transition
scenarios is shown in Fig.~\ref{f3}.

Accretion of matter onto the surface of the critical configuration
$C_N$ will lead to a jumpwise transition to the configuration $C_Q
$, having a finite-size core consisting the deconfined quark
matter. This transition will be accompanied by an enormous release
of energy determined by the difference in binding energies of
these configurations:
\begin{equation}
E_{release}=E_{bind}(C_Q)-E_{bind}(C_N)=(M(C_N)-M(C_Q))c^2.
\label{eq3}
\end{equation}

\begin{figure}[pb]
\centerline{\psfig{file=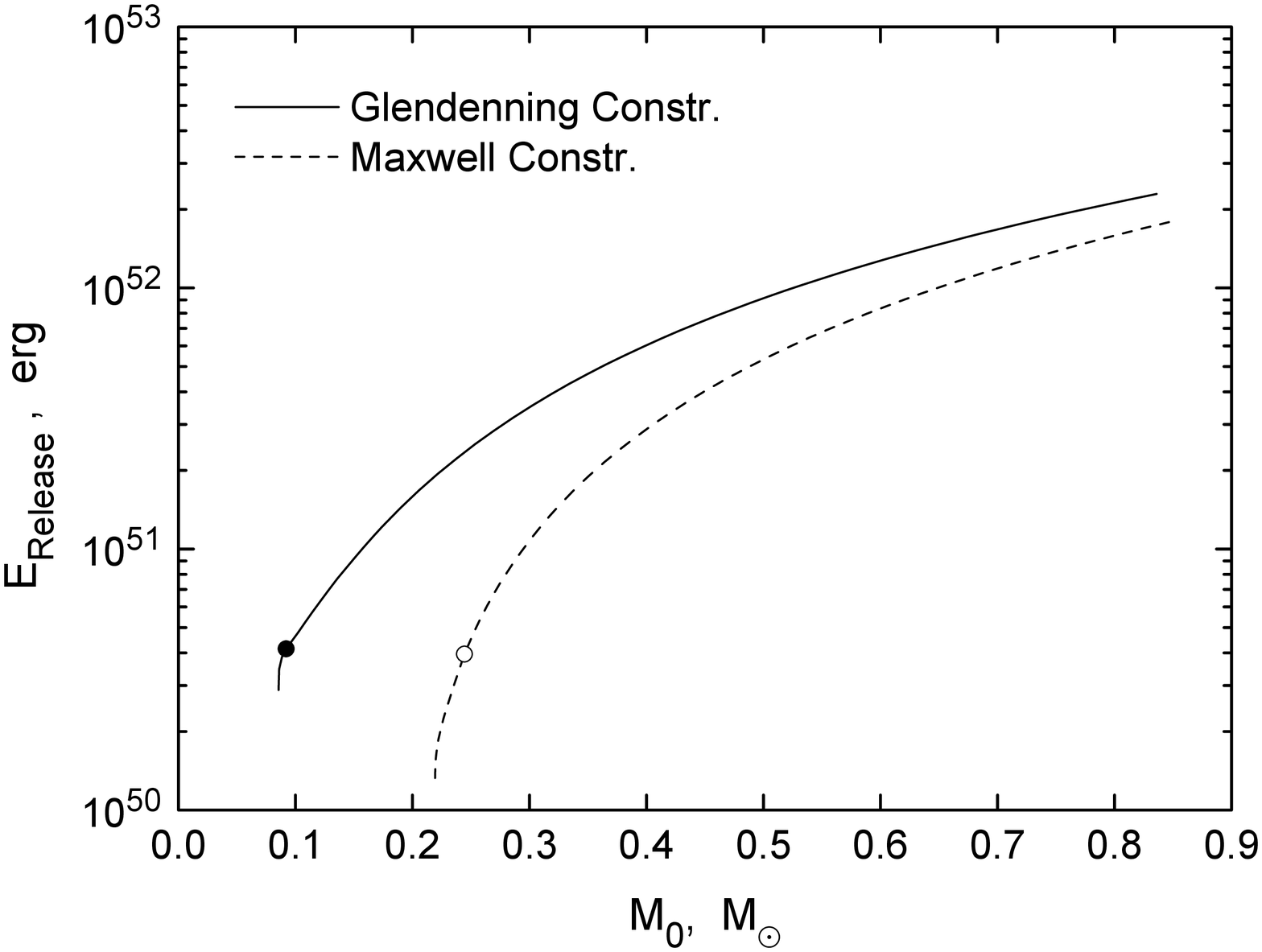,width=8 cm}} \vspace*{8pt}
\caption{The total energy release due to the hadron-quark phase
transition as a function of baryonic mass of neutron star $M_0$
for Maxwell and Glendenning constructions. Circles correspond to
the $C_N \rightarrow C_Q$ transition presented in Fig.~\ref{f3}.
\label{f4}}
\end{figure}
In Fig.~\ref{f4} we plot the total energy release as a function of
baryonic mass of star. One can see that in both cases of phase
transition constructions the released energy increases with the
increase of baryonic mass $M_0$. For a fixed value of the baryonic
mass $M_0$ of star the conversion energy  in the case of
Glendenning construction more than in the case of Maxwell
construction. In addition, the minimum required baryonic mass for
the catastrophic rearrangement of the neutron star and the
formation of a quark core in the center of the star greater in the
case of the Maxwell construction compared to the Glendenning
construction case. In the case considered here, the quark
deconfinement phase transition in the neutron star interior leads
to the energy release of the order $10^{50}\div 10^{52}$~erg.
\begin{figure}[pb]
\centerline{\psfig{file=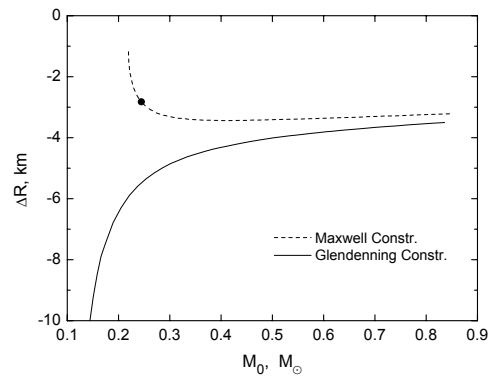,width=8 cm}} \vspace*{8pt}
\caption{Changes in the stellar radius due to the hadron-quark
phase transition as a  function of baryonic mass $M_0$ of a
neutron star for Maxwell and Glendenning scenarios. \label{f5}}
\end{figure}

Fig.~\ref{f5} shows the changes of stellar radius $\Delta
R=R_Q-R_N$ due to the quark deconfinement phase transition as a
function of baryonic mass $M_0$ of a compact star in both
Glendenning and Maxwell construction cases. It is seen that in
both cases the compact stars radii decrease. In addition, the
behavior of changes $\Delta R$ for the two alternative scenarios
are strongly distinguished.

\section{Conclusion}

In this article we have studied the changes of the internal
structure and observable characteristics of the near-critical
configurations of compact stars and associated energy release due
to the quark deconfinement phase transition. For description of
isospin-asymmetric nuclear matter we use the RMF theory, including
a scalar-isovector $\delta$-meson effective field. The quark phase
is described in frame of the improved version of the MIT bag
model. Using the obtained characteristics of hadronic and strange
quark matter phases, we calculate the neutron star matter EOS with
quark-deconfinement phase transitions, corresponding to the
Maxwell and Glendenning scenarios.

We find the dependence of conversion energy on the baryonic mass
of neutron stars and analyze the changes in stellar radiuses due
to the deconfinement phase transition.

We show that for a fixed value of the baryonic mass $M_0$ of star
the conversion energy  in the case of Glendenning construction
more than in the case of Maxwell construction. The minimum
required baryonic mass for the catastrophic rearrangement of the
neutron star and the formation of a quark core in the center of
the star greater in the case of the Maxwell construction compared
to the Glendenning construction case. In the case considered here,
the quark deconfinement phase transition in the neutron star
interior leads to the energy release of the order $10^{50}\div
10^{52}$~erg.

Our obtained results will give the opportunity to clarify the
observational differences between the two scenarios of
quark-deconfinement phase transition and to formulate a specific
test for determining the phase transition scenario taking place in
reality.

%\section*{Acknowledgments}

%\begin{thebibliography}{000} %for 3 digits

\end{document}